\documentclass[twocolumn]{emulateapj}

\usepackage{wasysym}
\usepackage{psfig}

\slugcomment{Published in ApJL}

\shorttitle{White dwarf habitable zone}
\shortauthors{Agol}

\begin{document}


\title{Transit surveys for Earths in the habitable zones of white dwarfs}


\author{Eric Agol}
\affil{Department of Astronomy, Box 351580, University of Washington,
    Seattle, WA 98195, USA; agol@uw.edu}
\affil{Kavli Institute for Theoretical Physics, University of California,
Santa Barbara, CA, 93106, USA}



\begin{abstract}
To date the search for habitable Earth-like planets has primarily focused
on nuclear burning stars.  I propose that this search should be expanded
to cool white dwarf stars that have
expended their nuclear fuel.  I define the continuously habitable zone
of white dwarfs, and show that it extends from $\approx$0.005 to 0.02
AU for white dwarfs with masses from 0.4 to 0.9 $M_\odot$, temperatures
less than $\approx 10^4$ K, and habitable durations of at least 3 Gyr.  As they are
similar in size to Earth, white dwarfs may be deeply eclipsed by
terrestrial planets that orbit edge-on, which can easily be detected with
ground-based telescopes.  If planets can migrate inward or reform near white
dwarfs, I show that a global robotic telescope network
could carry out a transit survey of nearby white dwarfs placing interesting
constraints on the presence of habitable Earths.  If planets were detected,
I show that the survey would favor detection of planets similar to Earth:
similar size, temperature, rotation period, and host star temperatures similar
to the Sun.  The Large Synoptic Survey Telescope could place even tighter 
constraints on the frequency of habitable Earths around white dwarfs.
The confirmation and characterization of these planets might be carried out with 
large ground and space telescopes.
\end{abstract}


\keywords{astrobiology --- binaries:eclipsing --- eclipses --- planetary systems --- planets and satellites:detection --- white dwarfs}



\section{Introduction}

The search for habitable planets has focused on stars similar to the
Sun as it is the sole example we have of a star with a habitable planet, and nuclear
burning provides a long-lived source of energy \citep{Kasting1993,Lunine2008}.
White dwarfs, which are as common as Sun-like stars, may also provide a
source of energy for planets for gigayear (Gyr) durations.
White dwarfs have typical masses 0.4---0.9 $M_\odot$ \citep{Provencal1998}, but have 
radii only $\approx$1\% of the
Sun, about the same size as the Earth \citep{Hansen2004}.
The most common white dwarfs have surface temperatures of $\approx$5000 K
which are referred to as ``cool white dwarfs''
since hotter white dwarfs are easier to detect \citep{Hansen2004}.
Cool white dwarfs typically have luminosities of $10^{-4}L_\odot$, so a planet
must orbit at $\approx$0.01 AU to be at a temperature for liquid water to exist
on the surface, the so-called habitable zone \citep{Kasting1993}.
The small size of white dwarfs can cause large transit depths by
Earth-sized or even smaller bodies which could in principle be detectable with 
ground-based telescopes \citep{DiStefano2010,Faedi2010,Drake2010}.

Prior to becoming a white dwarf, a Sun-like star expands to a red giant,
engulfing planets within $\approx$1 AU \citep[e.g.,][]{Nordhaus2010}.
Planets present in the white dwarf habitable zone (WDHZ)
must arrive after this phase.  This may occur via several 
paths \citep{Faedi2010}: planets can form out of gas near the white dwarf,
via the interaction or merger of binary stars \citep{Livio2005},
or by capture or migration from larger distances \citep{Debes2002}.
There are precedents for each of these processes: one neutron star has a
planetary system \citep{Wolszczan1992} which may have been formed from
a disk created after the supernova 
\citep{Phinney1993};  pulsars show low-mass stellar
companions being whittled down to planet masses 
\citep{Fruchter1988}; and white dwarfs show infrared emission and
atmospheric compositions indicative of close orbiting dust and/or bodies
\citep[e.g.,][]{Zuckerman2010}.
Consequently, short period planets around white dwarfs might plausibly
exist.  I sidestep the question of formation, which
has had little theoretical attention, and instead address 
the location and duration of habitable
zones around white dwarfs (Section \ref{wdhz}), the detection of planets in
the WDHZ, even if their frequency were much less than 1\% 
(Sections \ref{detection} and
\ref{survey}), and how characterization of these planets might proceed 
(Section \ref{characterization}).

\section{White dwarf habitable zone}  \label{wdhz}

I compute the WDHZ boundary following the procedure in \citet{Selsis2007}.  
I determine the white dwarf luminosity and effective temperature,
$T_{\rm eff}$, versus age from white dwarf cooling tracks \citep{Bergeron2001}.
With the luminosity and effective temperature, I compute the range of
distances within which an Earth-like planet could have liquid water
on the surface if it were placed there with an intact atmosphere.  I use 
computations of the limits of the habitable zone for stars of different 
effective temperatures that are based on empirical limits
from our solar system combined with one-dimensional radiative---convective atmospheric
models for Earth-like planets that include water loss at the inner edge and
the maximum CO$_{\rm 2}$ greenhouse effect at the outer edge \citep{Kasting1993}.

Most white dwarfs have masses $M_{\rm WD}=0.6 M_\odot$ and CO interior
composition, for which I plot the WDHZ versus time in Figure 1 as
a blue shaded region.  This region shrinks with time as the star
cools.  A planet enters at the bottom of Figure 1 and moves vertically
up the figure as its white dwarf host ages, so it starts off too hot for
liquid water, passes through the WDHZ, and then becomes too cold.
The duration a planet spends within the WDHZ, $t_{\rm HZ}$, has a maximum of
8 Gyr at $\approx$0.01 AU.  Based on the WDHZ limits, I next define the 
``continuously habitable zone'' (CHZ)
as the range of planet orbital distances, $a$, that are habitable for
a minimum duration, $t_{\rm min}$ (Figure 2).  I choose a minimum duration of 
$t_{\rm min} = 3$ Gyr, which results in a CHZ within $a < 0.02$ AU:  a planet 
that orbits within this distance spends at least 3 Gyr within the WDHZ.
From 0.4 to 0.9 $M_\odot$ with $t_{\rm min}$=3 Gyr the outer boundary of the
CHZ always falls within 10\% of 0.02 AU for hydrogen and helium atmospheres
(Figure 2).  To check the sensitivity to the white dwarf cooling computations,
I have also computed the WDHZ with BASTI models \citep{Salaris2010}, for
which I find a slightly longer $t_{\rm HZ}$.  I have also computed
the CHZ for atmospheres with $t_{\rm min}=$ 1 Gyr and
5 Gyr which shifts the outer boundary of the CHZ outward/inward by a factor of
$\approx$1.5/0.7 (Figure 2).

\begin{figure}
\centerline{\psfig{figure=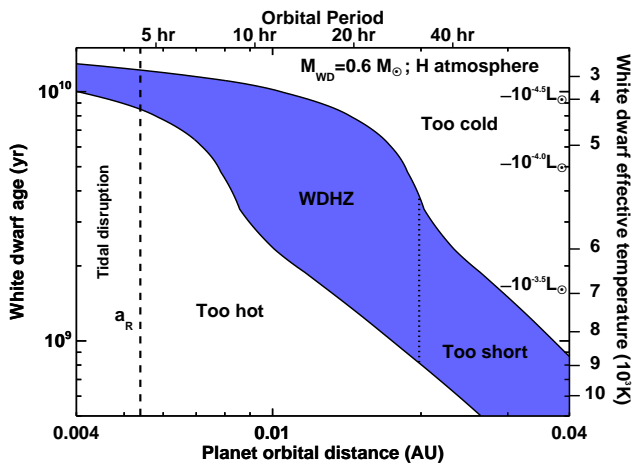,width=\hsize}}
\caption{WDHZ for $M_{\rm WD}=0.6 M_\odot$ vs.\ white dwarf age and
planet orbital distance.  Blue region denotes the WDHZ.  Dashed line is
Roche limit for Earth-density planets.  Planets to right of
dotted line are in the WDHZ for less than 3 Gyr.
Planet orbital period is indicated on the top axis; and white dwarf effective
temperature on the right axis.  Luminosity of the white dwarf at different ages
are indicated on right.}\label{fig01}
\end{figure}

\begin{figure}
\centerline{\psfig{figure=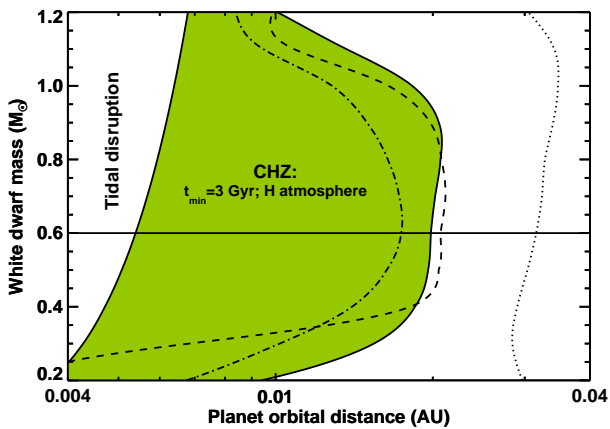,width=\hsize}}
\caption{CHZ vs.\ white dwarf mass and planet orbital distance.
Green region is the CHZ for $t_{\rm HZ} > t_{\rm min}$=3 Gyr, H-atmosphere.
Left solid line is Roche limit for Earth-density
planets.  The other lines show how the CHZ outer boundary changes
for $t_{\rm min}$=1 Gyr (dotted line), for $t_{\rm min}$=5 Gyr (dash-dotted line), or
for an He atmosphere with $t_{\rm min}$= 3 Gyr (dashed line).  Horizontal line
indicates the most common white dwarf mass of 0.6 $M_\odot$, plotted in Figure 1.}
\end{figure}

There are several important consequences of the WDHZ and CHZ.
First, the range of white dwarf temperatures in the portion of
the CHZ within the WDHZ is that of cool white dwarfs, 
$\approx$3000--9000 K (right hand axis in Figure 1), similar
to the Sun.  At the hotter end higher ultraviolet
flux might affect the retention of an atmosphere, these planets would
need to form a secondary atmosphere, as occurred on Earth.  Excluding higher
temperature white dwarfs only slightly modifies the CHZ since they spend 
little time at high temperature.  Cool white dwarfs are photometrically stable
\citep{Fontaine2008},
which is critical for finding planets around them.  Second, for white 
dwarfs with temperatures $\ga$4500 K, the WDHZ is exterior to the Roche
limit (tidal disruption radius),
$a_{\rm R} = 0.0054 {\rm AU} (\rho_{\rm p}/\rho_\oplus)^{-1/3}
(M_{\rm WD}/0.6 M_\odot)^{1/3}$,
where $\rho_{\rm p,\oplus}$ are the mass densities of the planet and of Earth.
Consequently, the 3 Gyr CHZ lies between $a_{\rm R}<a<0.02$ AU,
indicated with the green region in Figure 2.

Third, the CHZ occurs at white dwarf luminosities of $10^{-4.5}$ to 
$10^{-3} L_\odot$,
about 10 magnitudes fainter than the Sun, which sets the minimum telescope
size for detection.  Finally, the orbital period of white dwarfs in the CHZ
is $\approx$4--32 hr.
At this period the timescales for tidal circularization and tidal locking
are $\approx$10--1000 years, so rocky planets will be synchronized and 
circularized \citep{Heller2011}; the side of the planet near the star will have a permanent 
day, while the far side will have a permanent night.  The planet will
orbit stably as it cannot raise a tide on the compact white dwarf.
Radiation drag will not cause the orbit to decay,
but magnetic field drag might, depending
on the planet's conductivity \citep{Li1998}.

\section{Detection of planets in the white dwarf habitable zone} \label{detection}

I plot example light curves of planets in the WDHZ in Figure 3 for
sizes within a factor of $\approx$2 of Earth orbiting a white dwarf
near the peak of the white dwarf luminosity function.
The transits last $\approx$2 minutes, and have a maximum depth of
10\%---100\%;  these events can be detected at a 100 pc distance with a 1 m
ground-based telescope.
To detect planets in the CHZ one must monitor a sample of white dwarfs for the
duration of the orbital period at 0.02 AU, and planets present in edge-on orbits will be
seen to transit their host stars.
The transit probability is
\begin{equation}
p_{\rm trans}= 1.0\% \left(\frac{R_{\rm p}/R_\oplus+R_{\rm WD}/0.013R_\odot} {a/0.01 {\rm AU}}\right),
\end{equation}
so for every 100 Earths orbiting white dwarfs at 0.01 AU with random orientations, 
on average one will be seen to transit.

\begin{figure}
\centerline{\psfig{figure=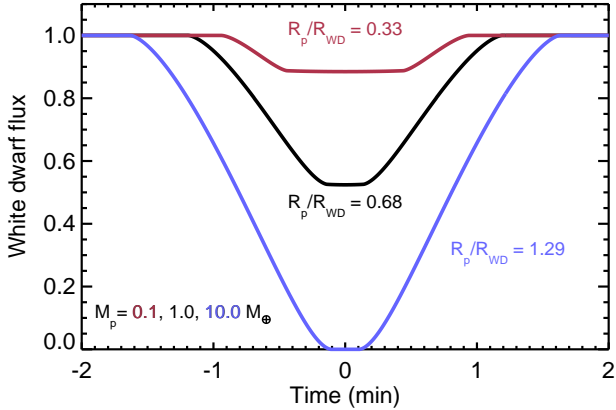,width=\hsize}}
\caption{Example light curves of habitable Earth-like planets transiting
a 0.6 $M_\odot$ white dwarf ($T_{\rm eff}=5200$K, $R_{\rm WD}=0.013 R_\odot$,
inclination $= 89^\circ.9$, 10\% linear limb darkening, and $a=0.013$ AU)
with masses of $0.1 M_\oplus$ (red, about Mars mass), $1.0
M_\oplus$ (black, Earth twin), and $10.0 M_\oplus$ (blue, super-Earth).
The ratios of the planet radii, $R_{\rm p}$, to white dwarf radius, $R_{\rm WD}$,
are indicated.}
\end{figure}

I define $\eta_\oplus$ to be the number of planets with $0.1 M_\oplus < M_{\rm p} <
10 M_\oplus$ in the 3 Gyr CHZ ($a<$ 0.02 AU).  To measure $\eta_\oplus$ to
an accuracy of $\approx$33\% requires detecting $\approx$9 planets, so one must survey 
$\approx$10$^3 \eta_\oplus^{-1}$ white dwarfs.

\section{Survey strategy} \label{survey}

The local density of white dwarfs is $(4.7\pm 0.5) \times 10^{-3}$ pc$^{-3}$ 
\citep{Harris2006}.
For a survey out to $D_{\rm max} < 200$ pc, the white dwarfs should be nearly
isotropically spaced on the sky at one per $\approx$2$(100 {\rm pc}/D_{\rm max})^{3}$ deg$^2$.
This exceeds the field of view of most telescopes, so each white dwarf must
be separately surveyed for the presence of transiting planets.

I have simulated an all-sky survey with a worldwide network of 1 m aperture
telescopes to monitor the white dwarf CHZ (typically 32 hr, during
which telescopes distributed in longitude follow a single star) following 
\citet{Nutzman2008} to compute the telescope sensitivity, including
sky and read noise, and assuming an exposure time of 15 s; 
I conservatively expanded the error bars an additional 50\%.
Each white dwarf is given a multi-planet system whose innermost planet is
chosen from a log-normal centered at $2a_{\rm R}$ with width of 0.5 dex, with
subsequent planets packed as closely
as allowed by dynamical stability on a timescale of $10^9$ yr \citep{Zhou2007},
leading to a uniform distribution in $\log{a}$, with a gradual cutoff within
$2a_{\rm R}$.  The planet masses are drawn from $dn/dM_{\rm p} \propto M_{\rm p}^{-\alpha}$ from
$10^{-2}M_\oplus$ ($\approx$Moon) to $10^2M_\oplus$ ($\approx$Saturn),
with $\alpha=4/3$ to match
the observed slope measured with the {\it Kepler} satellite \citep{Borucki2011},
as well as that found in the solar system.  I assumed Earth-composition
planets with the radius determined from the mass according to
\citet{Seager2007}.  Two exposures within transit 
with a signal-to-noise of at least 6 constitute a detection;  this keeps the 
false-positive level to $<$1\% for the entire survey.  I then scale the
simulated detection rates with $\eta_\oplus$ to determine the expected
number of detected planets.

To detect 9$\pm$3 planets, for $\eta_\oplus$ = 50\% a survey of 2800 white
dwarfs within 52 pc is required for a total on-sky time of 10 years (for
a single telescope with 31 hours per white dwarf on average). A smaller
planet frequency of $\eta_\oplus$ = 10\% requires $\approx$20,000 white
dwarfs within 100 pc for a total 69 years of telescope time on sky.
For a network of twenty 1 m telescopes distributed
around the globe, such as the Las Cumbres Observatory Global Telescope
\citep[LCOGT;][]{Hidas2008}, or the Whole Earth Telescope 
\citep[WET][]{Nather1990}, devoted
to observing white dwarfs at 25\% efficiency (50\% of time at night,
50\% weather loss), the total calendar time required would be 2 years
for a survey of 2800 white dwarfs and 14 years for a survey of 20,000.
If the CHZ is surveyed to only 0.01 AU, this would decrease
the required calendar time to 8.5 months and 5 years, respectively, but would 
also decrease the planet yield.

Figure 4 shows the probability distribution for planets detected in 10$^4$ survey
simulations. Each one surveys 20,000 single white dwarfs out to 100 pc
for planets within 0.02 AU.
Of the detected planets, an average of 40\% will be currently within the WDHZ.
Remarkably, the detection probability peaks near planets of the size and 
temperature of Earth due to the 
coincidence in size of the Earth and white dwarfs, and the coincidence between 
the WDHZ at the peak of the white dwarf luminosity function and $2a_{\rm R}$.
This leads to the following biases: (1) large planets have a higher transit 
detection probability $\propto R_{\rm p} dn/dR_{\rm p}$, so the number
detected declines if $dn/dR_{\rm p}$ is steeper than $R_{\rm p}^{-1}$;
(2) small planets cause shallower transits for $R_{\rm p} \ll R_{\rm WD} 
\approx R_\oplus$, so the detection rate scales as 
$\propto R_{\rm p}^6 dn/dR_{\rm p}$ \citep{Pepper2003}, although
the range of luminosities of white dwarfs flattens this decline;
(3) cooler planets have a smaller probability of transit, so fewer are detected as
$\propto T_{\rm p}^4$ for small $T_{\rm p}$ and
a uniform distribution in $\log{a}$; and (4) hotter planets orbit hotter
stars, which are less numerous, $dn/dT_{\rm WD} \propto T_{\rm WD}^{-3.9}$
for large $T_{\rm WD}$ \citep{Hansen2004}.  Although these trends should occur 
for any volume-limited survey, the
break in the radius detection limit depends on the size of the
telescope and signal-to-noise cutoff that is chosen:  larger telescopes
or smaller signal-to-noise cuts will be more sensitive to small radius planets.
Figure 4 is sensitive to the
properties of the planet population: if the inner cutoff, $a_{\rm in}$, is 
further/closer, then the peak temperature moves to
cooler/hotter temperatures, $\propto a_{\rm in}^{-1/2}$, and the total number of 
planets detected declines/increases as $a_{\rm in}^{-1}$ due to the lower/higher transit 
probability.  If the planet size distribution is steeper/flatter, then the detected 
planet distribution peaks at smaller/larger sizes.

\begin{figure}
\centerline{\psfig{figure=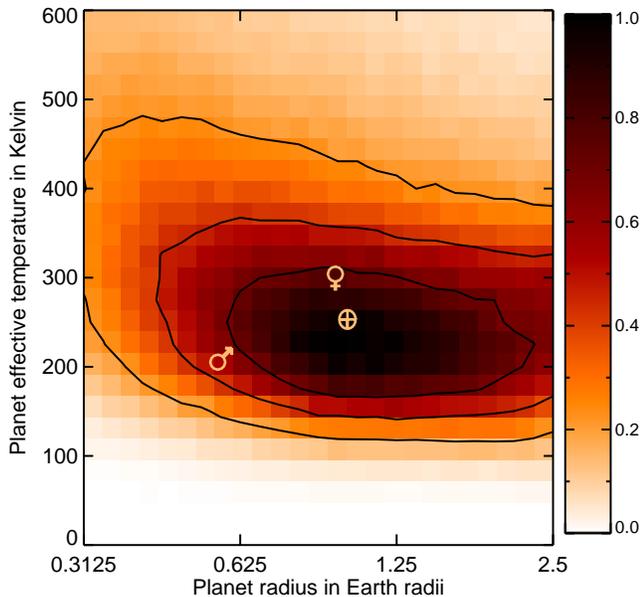,width=\hsize}}
\caption{Probability density, $d^2n/(dT_{\rm p}d\log{R_{\rm p}})$, of detected planets
vs.\ planet radius, $R_{\rm p}$ (log axis scale), and planet effective temperature, $T_{\rm p}$ 
(assuming the same albedo as Earth).  The contours enclose 25\%, 50\%, and 75\% of all 
detected planets; the contour levels are 29\%, 53\%, and 76\% of the peak density.
Earth ($\oplus$), Mars ($\mars$), and Venus ($\venus$) symbols indicate the radii 
and effective temperatures of these solar system planets.}
\end{figure}

Prior to such a survey, a nearby sample of cool white dwarfs must be found using measurements
of the reduced proper motion.  Ongoing and planned deep astrometric surveys, such as
90 Prime, Skymapper, Pan-STARRS, URAT, and GAIA, should find most cool
white dwarfs out to 100 pc ($V$$<$21) within the decade \citep{Henry2009,
Kalirai2009}.  Some of these surveys might also find transits
if the requirement of three epochs in transit is relaxed.
For example, GAIA will observe 200,000 disk white dwarfs 50-100 times each
\citep{Perryman2001}, possibly detecting one epoch in eclipse for $\approx$10\% 
of these stars with habitable transiting planets.

For values of $\eta_\oplus$$<$10\%, more stars must be observed to detect
planets, so a better strategy is to observe multiple white dwarfs simultaneously
with a wide-field imager with fast readout, such as the Large Synoptic Survey Telescope \citep[LSST][]{LSST2009}.
The LSST survey is expected to detect $\approx$10$^7$ white dwarfs over half
of the sky at $>$5$\sigma$ to $r$$<$24.5 with $\approx$1000 epochs each and two
15 s exposures per epoch over the duration of the 10 year survey.  Since
LSST is a magnitude-limited survey, the white dwarf temperature distribution peaks
at $10^4$K.  I have taken the simulated detected
distribution of white dwarfs for LSST \citep{Juric2008},
created simulated light curves for white dwarfs with planets, and added
noise \citep{LSST2009}.  I find LSST can detect $>$9 CHZ planets if
$\eta_\oplus$$>$5$\times$10$^{-3}$, where detection requires
that at least three epochs fall within transit with two points each
detected at $>$7$\sigma$.  The LSST survey will be biased toward detecting
shorter period ($\propto P^{-4/3}$) and large-size planets that have
yet to enter the WDHZ since their stars are hotter.  This could be improved by
either continuously observing some fields for several nights, or taking more exposures 
per field, resulting in detection of smaller, cooler planets, thus constraining smaller 
$\eta_\oplus$.  LSST will identify the white dwarfs with reduced proper motion 
measurements as the survey is being carried out.  

I estimate that $>$10$^3$ double white 
dwarf eclipsing binaries with orbital periods similar to WDHZ planets will be found with
LSST using the BSE population synthesis model for binary stars \citep{Hurley2002} with
parameters taken from observed binaries \citep{Raghavan2010}.  I find that about
2.5\% of white dwarfs will have a white dwarf companion with a period in the 
range of 8---64 hr which might be mistaken for a transiting CHZ planet if 
these are viewed edge-on.  Follow-up of planet candidates will be required to 
distinguish the two possibilities: 
white dwarf binaries will show primary and secondary eclipses of different depths 
(if the two white dwarfs differ in
temperature), offset secondary eclipses due to light travel time,
gravitational lensing \citep{Agol2002}, and Doppler modulation, and 
an eclipse shape that differs from planetary transits if non-grazing.  I 
simulated light curves of white dwarf binaries including these effects and find 
that in the worst case of two white dwarfs with identical temperatures
these distinguishing features may be detected with 10---100 m
ground-based telescopes for systems out to 50---100 pc, either photometrically
or with radial velocities. 
Another concern is grazing eclipses from white dwarf/M dwarf binaries;  
these can be identified by the eclipse shapes, spectral energy distribution, and differing secondary eclipse depths.

\section{Planet characterization} \label{characterization}

The parallax and spectrum of a white dwarf yield its mass, luminosity, atmospheric
composition, and radius; then the transit depth gives the planet radius.
The planet's mass cannot be measured from Doppler shifts
due to the featureless spectra of cool white dwarfs,
but may be bracketed by the range
of compositions for planets of a given size \citep{Seager2007}.
The mass might be measured
by observing wavelength dependent absorption, such as Rayleigh scattering
that varies as $8H/R_{\rm p} \ln{\lambda}$ \citep{LecavelierDesEtangs2008},
where $H$ is the atmospheric scale height, causing the transit depth
to vary by $\approx$few millimagnitude over 400---500 nm if Rayleigh scattering dominates
over other sources of opacity.  If atmospheric molecular weight can be 
estimated, so can the planet mass \citep{MillerRicci2009}.
If two planets transit a white dwarf (about 2.5\% of
the time for mutual inclinations within $5^\circ$ and in a
packed planet system), then transit timing variations might
constrain the planet masses if the orbital period ratio is nearly commensurate 
\citep{Holman2005,Agol2005}.
Infrared phase variation of planets \citep{Knutson2007} in the CHZ is
$\approx$0.5\%---2\% at 15---20 $\mu$m; however, I estimate that the {\it James Webb
Space Telescope} ({\it JWST}) cannot detect this for cool white dwarfs due to
telescope noise.
For a hotter white dwarf of $10^4$ K at 100 pc with an Earth-like planet
at 0.01 AU with a day---night contrast of 30\%, the phase variation of 0.1\%
at 7.7 $\mu$m might be detectable at 9$\sigma$ with the MIRI imager on
{\it JWST} \citep{Swinyard2004}.

Several topics require further study.  The global climate models for
determining the WDHZ should include fast synchronous rotation, magnetic fields,
varied planet atmospheric composition, radiogenic heating, and white dwarf
cooling.  The WDHZ is a necessary but insufficient criteria for habitability.
For example, planets that start hot may not retain their atmospheres, as has 
also been argued for planets orbiting M dwarfs \citep{Lissauer2007};  this may
require volatile delivery from more distant bodies in the system 
\citep{Jura2010} or planetary outgassing.  To retain an atmosphere
might require a larger planet escape velocity, possibly favoring super-Earths
for habitability. 

Formation mechanisms must be modeled to help motivate future
surveys.  For example, gravitational interactions
of a planet and star with a third companion body may be responsible for 
creating hot Jupiters \citep{Fabrycky2007}, which is also promising 
for moving 
distant planets around white dwarfs to $2a_{\rm R} \approx 0.01$ AU, the tidal
circularization radius \citep{Ford2006}.  
It is also possible that tidal
disruption of a planet or a companion star will result in the formation of 
a disk which may cool and form planets \citep{Guillochon2010}, out of which 
a second generation of planets might form \citep{Menou2001,
Perets2010,Hansen2009}.  

The most common white dwarf has $T_{\rm eff}\approx$5000 K, close to that
of the Sun;  consequently, inhabitants of a planet in the CHZ will see their 
star as a similar angular size and color as we see our Sun.
The orbital and spin period of planets in the CHZ are similar
to a day, causing Coriolis and thermal forces similar to Earth.
The night sides of these planets
will be warmed by advection of heat from their day sides if
a cold-trap is avoided \citep{Merlis2010}.
Transit probabilities of habitable planets are similar for cool white
dwarfs and Sun-like stars, but the white dwarf planets can be found using
ground-based telescopes (e.g., LCOGT, WET, and  LSST) at a much less expensive price
than space-based planet-survey telescopes.





\acknowledgments
I acknowledge NSF CAREER grant AST-0645416, and thank KITP
(NSF PHY05-51164) and the Whiteley Center for their hospitality.
I thank Fred Adams, Rory
Barnes, Pierre Bergeron, Tim Brown, Mark Claire, Nick Cowan, Dan
Fabrycky, Eric Gaidos, Rob Gibson, Brad Hansen, Mario Juric, Lisa
Kaltenegger, Piotr Kowalski, Jim Kasting, Vikki Meadows, Enric Palle,
Michael Perryman, David Spiegel, Giovanna Tinetti, and the anonymous 
referee for help.

\noindent{\it Note added in proof.} Ren\'e Heller informed me that
\citet{Monteiro2010} discussed the white dwarf habitable zone for known
white dwarf stars, showing a boundary similar to that in Figure 1.

\end{document}